\documentclass[]{aastex63}

\usepackage{multirow}
\usepackage[T1]{fontenc}
\usepackage{amssymb}
\usepackage{graphicx}
\received{}
\revised{}
\accepted{}
\submitjournal{ApJ}

\shorttitle{Fast Transient Finding Algorithm}
\shortauthors{Strausbaugh et al.}
\graphicspath{{./}{figures/}}

\begin{document}

\title{Finding Fast Transients in Real Time Using Novel Light Curve Analysis Algorithm}

\correspondingauthor{Robert Strausbaugh}
\email{robert.strausbaugh@uvi.edu}

\author[0000-0001-6548-3777]{Robert Strausbaugh}
\affiliation{University of the Virgin Islands \\
Number 2 Brewers Bay Rd. \\
St. Thomas, VI 00802, USA}

\author{Antonino Cucchiara}
\affiliation{College of Marin}
\affiliation{University of the Virgin Islands \\
Number 2 Brewers Bay Rd. \\
St. Thomas, VI 00802, USA}

\author{Michael Dow Jr.}
\affiliation{University of the Virgin Islands \\
Number 2 Brewers Bay Rd. \\
St. Thomas, VI 00802, USA}

\author{Sara Webb}
\affiliation{Centre for Astrophysics and Supercomputing, Swinburne University of Technology, Mail Number H29, PO Box 218, 31122 Hawthorn, VIC, Australia}

\author[0000-0001-5310-4186]{Jielai Zhang}
\affiliation{Centre for Astrophysics and Supercomputing, Swinburne University of Technology, Mail Number H29, PO Box 218, 31122 Hawthorn, VIC, Australia}
\affiliation{ARC Centre of Excellence for Gravitational Wave Discovery (OzGrav), Hawthorn, 3122, Australia}

\author{Simon Goode}
\affiliation{Centre for Astrophysics and Supercomputing, Swinburne University of Technology, Mail Number H29, PO Box 218, 31122 Hawthorn, VIC, Australia}

\author{Jeff Cooke}
\affiliation{Centre for Astrophysics and Supercomputing, Swinburne University of Technology, Mail Number H29, PO Box 218, 31122 Hawthorn, VIC, Australia}

\begin{abstract}
    The current data acquisition rate of astronomical transient surveys and the promise for significantly higher rates in the next decade necessitate the development of novel approaches to analyze astronomical data sets and promptly detect objects of interest.  The Deeper, Wider, Faster (DWF) program is a survey focused on the identification of fast evolving transients, such as fast radio bursts, gamma-ray bursts, and supernova shock breakouts.  
    It employs a multi-frequency  simultaneous coverage of the same part of the sky over several orders of magnitude. Using the Dark Energy Camera mounted on the  4-meter Blanco telescope, DWF captures a 20 second g-band exposure every minute, at a typical seeing of $\sim 1''$ and an airmass of $\sim1.5$.  These optical data are collected simultaneously with observations conducted over the entire electromagnetic spectrum -- from radio to $\gamma$-rays -- as well as cosmic ray observations.  In this paper, we present a novel real-time light curve analysis algorithm, designed to detect transients in the DWF optical data; this algorithm functions independently from, or in conjunction with, image subtraction. We present a sample of fast transients detected by our algorithm, as well as a false-positive analysis.  Our algorithm is customizable and can be tuned to be sensitive to transients evolving over different timescales and flux ranges.
    
\end{abstract}



\section{Introduction} \label{intro}
The field of transient astronomy is booming, with several successful completed, ongoing, and planned optical surveys  that will come online in the coming years, specifically designed to find transient phenomena.  Among the former, the Palomar Transient Factory/Intermediate Palomar Transient Factory \citep[PTF/iPTF,][]{ptf,ptf2} 
the Panoramic Survey Telescope and Rapid Response System \citep[Pan-STARRS,][]{panstarrs}, 
the Sloan Digital Sky Survey (SDSS) Supernova Survey \citep[][]{sdss_snsurvey},
the Asteroid Terrestrial-impact Last Alert System (ATLAS) All-sky Survey \citep{atlas_survey}, 
the Gaia Survey \citep{gaia_survey},
the Zwicky Transient Facility \citep[ZTF,][]{ztf,ztf_science} 
the Dark Energy Survey \citep[DES,][]{des} 
the All Sky Automated Survey for Supernovae \citep[ASAS-SN,][]{asassn} 
and the Transiting Exoplanet Survey Satellite \citep[TESS,][]{tess} 
have provided a census of a large variety of supernovae (SNe), 
tidal-disruption events, and exo-planet confirmations. In the latter category, the Vera C.\ Rubin Observatory 
and Nancy Grace Roman Telescope 
will push our understanding of the transient sky towards deeper limits and longer wavelengths.
An overview of the field of view (FoV), depth, and cadence of these surveys can be found in Table \ref{survey_table}.  In conjunction with some of these optical surveys, gravitational wave (GW) detectors like the Advanced Laser Interferometer Gravitational-wave Observatory \citep[aLIGO,][]{aligo} and Virgo \citep{virgo}, and neutrino detectors such as IceCube \citep{icecube} and the Baksan Neutrino Observatory \citep{bno} have ushered us in a new era of multi-messenger transient astronomy.  

\begin{deluxetable*}{cccc}
\tablenum{1}
\tablecaption{Optical Transient Survey Details\label{survey_table}}
\tablewidth{0pt}
\tablehead{
&\colhead{Field of View}\vspace{-0.4cm}&&\colhead{Cadence [days unless}\\
\colhead{Survey Name}&\vspace{-0.4cm}&\colhead{Depth}&\\
&\colhead{[square degrees]}&&\colhead{otherwise noted]}
}
\startdata
Pan-STARRS1 & 7 & $g\sim22$ & 90$^a$ \\
SDSS SN Survey & 3 & $r\sim22$ & 4 \\
PTF/iPTF SN Survey & 7.8 & $g\sim 21.3$ & 5 \\
ATLAS & 28.9 & $m\sim19$ & 2\\
Gaia & 2 $\times$ 0.5 & G$\sim20.7$ & 20$^a$\\
DES & 2.5 & $g\sim25$ & 5 \\
ZTF & 47 & $g\sim20.5$ & 1-3 \\
ASAS-SN & 4.5 & $v\sim17$ & 2-3\\
TESS & 4 $\times$ 24 & broadband$^b$ $\sim15$ & 2-30 minutes\\
Rubin & 9.6 & $g\sim24$ & 3$^a$\\
Roman & 9 & $J\sim$25 & 5\\
\textbf{DWF} & $2.5^{c}$ & \textbf{$g\sim24$} & \textbf{1 minute}
\enddata
\tablecomments{The field of view, depth, and cadence of notable past, present, and future transient surveys. We note that surveys with $^a$ have or will at some point operate at cadences shorter than those listed: e.g., the Deep Drilling fields for Rubin \citep{lsst_deepdrilling} and fast cadence search in Gaia \citep[][]{gaia_fast} The TESS broadband$^b$ wavelengths span $R_c$-, $I_z$-, and $z$-bands. $^c$ The field-of-view quoted for DWF is the ``on-sky'' area of the science CCDs and not the footprint FoV that includes CCD gaps that is typically listed.}
\end{deluxetable*}

We expect that the discovery of new and exciting transient phenomena will continue at higher pace thanks to the Vera C.\ Rubin Observatory/Legacy Survey of Space and Time \citep[LSST,][]{lsst}: 
the telescope is being constructed in Cerro Pachón, Chile (with a planned first light date in 2022 before commencing operations in 2023), and aims to image the sky in Wide, Fast, Deep mode at a depth of $g\sim24$ every 3 days \citep{lsst_manual}.  The Nancy Grace Roman Telescope/Wide-field Infrared Space Telescope \citep[WFIRST,][]{wfirst} will cover an area of 9 square degrees at an average depth of $J\sim25$ with a cadence of 5 days in a proposed medium depth supernova survey when it is launched on its scheduled date in 2025 \citep{wfirst_manual,wfirst_snsurvey}. 
We summarize the characteristics of these planned instruments in Table \ref{survey_table}.

\subsection{Current Survey Results}

In this section we briefly summarize some of the seminal discoveries that the next-generation of surveys will look to build upon.  For example, Pan-STARRS results include the outburst from a SN progenitor one year before its explosion \citep{panstarrs1}, a lack of super-luminous SNe with light-curves compatible with pair-instability models \citep{panstarrs2}, 
and the first interstellar asteroid detection \citep{interstellar_asteroid}.  Similarly, the most recent SDSS survey results include the detection of baryonic acoustic oscillation measurements \citep{sdss1}, evidence for the Epoch of Reionization around $z\sim6$ \citep{sdss2}, high redshift ($z>5.6$) quasars \citep{sdss3}, and indirect dark matter detection via weak-lensing \citep{sdss4}.  And finally, in the recent years PTF/iPTF/ZTF have further enhanced our knowledge on tidal disruptions events \citep[TDEs,][]{ptf_tde,ztf_tde}, gamma-ray burst (GRB) orphan afterglows \citep{ptf_orphanafterglow,ztf_orphanafterglowcandidate}, and  hosts of novel SNe \citep{ptf_sn1,ptf_sn2,ptf_sn3,ptf_sn4,ptf_sn5,ptf_sn6,ptf_sn7}.

\begin{figure}
    \centering
    \scalebox{0.35}[0.35]{\includegraphics{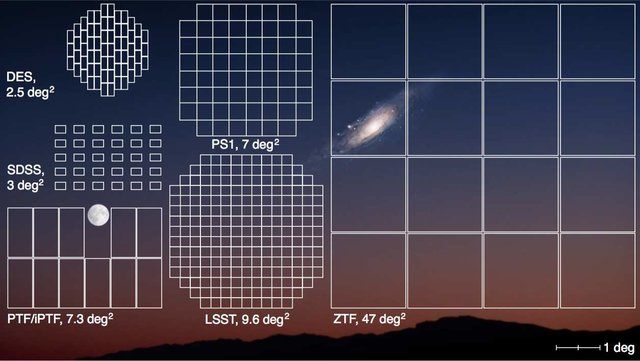}}
    \scalebox{0.35}[0.35]{\includegraphics{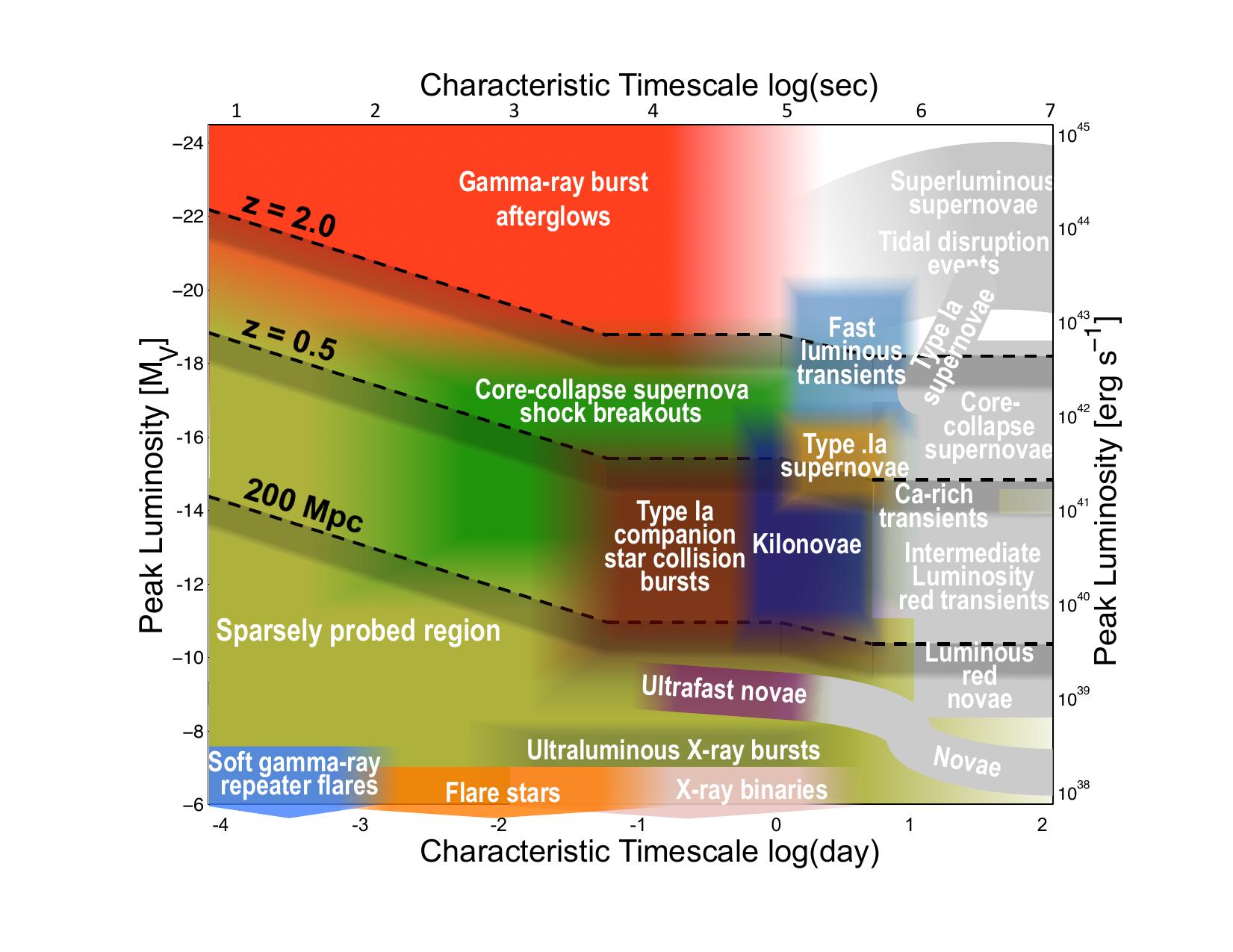}}
    \caption{Left: Adopted from \cite{survey_fov}.
    Comparison of the field of views of various
    astronomical surveys.  The DWF program uses the same telescope as the DES, the Blanco 4-meter scope at Cerro Tololo Inter-American Observatory in Chile.  Right: The characteristic timescales and brightness of transients within the DWF detection phase space.  Adopted from Cooke et al. (2022, in prep.).}
    \label{counthistograms}
\end{figure}

Due to the design of these facilities and their observational strategies, there is still a lack of discoveries of fast ($\lesssim$ 1 hour) and faint ($m_g \gtrsim 21$) transients (see Figure~\ref{counthistograms} right panel).  There are several science cases that benefit from the fast identification (and potential classification) of rapidly varying transients
with simultaneous observations spanning the electromagnetic spectrum.  Fast radio bursts (FRBs) are a class of objects 
characterized by either single, or repeating radio bursts evolving on sub-second time scales; finding emission at frequencies other than radio is important toward understanding the progenitors and ignition mechanism(s) for these events.  The first emission other than radio observed to accompany an FRB was recently detected from a magnetar in the Milky Way at X-ray frequencies; although the bursts are weaker in the radio than extragalactic FRBs, this event suggests that magnetars may be the progenitors of at least some FRBs \citep{frb_em1,frb_em2,frb_em3}.  

Also, due to a lack of sufficient early time observations of SNe, the shock breakout mechanism for type Ia SNe
and the ignition mechanisms for core-collapse SNe (CC-SNe) are still not well understood. Type Ia SNe
are critical for determining cosmological distance scales, and are used to measure the accelerated
expansion of the universe \citep{sn_accexp}. Despite their use as standard candles, the shock breakout mechanism
for type Ia SNe is not conclusively understood \citep{1a_detmech}.  Identification and follow-up of a type Ia SNe
within the first day, building on early-time observations of Type Ia Supernovae by ZTF \citep{ztf_early_ia} and TESS \citep{tess_early_ia}, can lead to an understanding of the shock breakout mechanism including the
detection of a cooling tail, indicative of a delayed detonation transition \citep{1a_shockbreakout}.  More importantly, the nature and progenitors of SNe Ia is unclear. The detection of UV/optical bursts from SNe Ia ejecta colliding with a companion star \citep[e.g.,][]{ia_progenitor} helps to secure the single-degenerate progenitor model for some fraction of events; these detections must occur on short timescales.  A better understanding of type Ia SNe could be the key towards resolving the more than $3\sigma$ discrepancy between measurements of the Hubble Constant using the cosmic microwave background \citep[e.g. by Planck,][]{cmb_H0,cmb_H0_2} and measurements made using the cosmic distance ladder method \citep[e.g.,][]{sn_accexp,cosmic_distance_ladder}, on which type Ia SNe are a vital rung.

The study of CC-SNe
is important in understanding the ends of the life cycles of massive stars and is believed to be one of the
drivers of nucleosynthesis of elements heavier than iron \citep{sn_nucleo}, in addition to collapsars \cite{collapsar_nucleo}. Early detection and follow-up of CC-SNe can distinguish between various theorized ignition mechanisms such as: magneto-rotational instabilities \citep{snbreakout_mri}, standing accretion shocks \citep{snbreakout_accretionshocks}, acoustic shocks \citep{snbreakout_accousticshock}, and QCD phase transitions \citep{snbreakout_qcd}.

The first GRB orphan afterglow may have been detected in the radio \citep{radio_orphanafterglow}, but was discovered at such a late time after its prompt emission, that deep optical follow ups resulted in only upper limits.  Searches are also being performed by ZTF \citep{ztf_orphan,ztf_orphanafterglowcandidate} for orphan afterglows and kilonovae.  One candidate orphan afterglow \citep{ztf_orphanafterglow1} was later associated with prompt $\gamma$- and X-ray emission from GRB 201103B \citep{ztf_orphanafterglow2}; the optical component was reported first by ZTF, instilling confidence in the veracity of their method in detecting and identifying orphan afterglows.  The study of orphan afterglows would allow us to calculate GRB jet angles, as well as the true GRB rate \citep{rhoads1997_grb}.

Although we have made many discoveries with surveys such as PTF, ZTF, and ASAS-SN, and plan on continued success with the Vera C.\ Rubin Observatory and Nancy Roman Space Telescope, there is a void in the parameter space for fast, faint transients that remains unfilled.  
Our understanding of GRB orphan afterglows, short GRBs, FRBs, SN ignition mechanisms and shock-breakouts, and electro-magnetic counterparts to gravitational wave events can be greatly enhanced by detecting these transients in real-time across several segments of the  electromagnetic  spectrum.  Due to the rarity of these events, the use of wide-field facilities is needed.

In this paper we present an automated, customizable fast transient identification algorithm centered mainly on Deeper Wider Faster program (DWF) source light curve analysis.  We summarize the DWF program in Section~\ref{sec:DWF} and describe the DWF data sets analyzed in this work.  In Section~\ref{algo} we motivate the need for a transient detection algorithm that is independent from image subtraction, and present the elements of a novel fast transient detection algorithm.  The results of running the algorithm on both real-time DWF data and later-time further processed data sets is presented in Section~\ref{results}. Finally in Section~\ref{conclusion}, we describe how this algorithm will be deployed in future DWF runs and how it can be used with data from other surveys.

\section{Deeper, Wider, Faster Program} \label{sec:DWF}

The Deeper, Wider, Faster program \citep[DWF,][]{dwf}, is peculiar among all the aforementioned transient surveys. The primary goal of DWF is to identify transient phenomena on the shortest timescales; DWF searches from milliseconds to hours in various wavelengths.  The deep optical component of DWF is carried out by the Dark Energy Camera \citep[DECam][]{decam} collecting 20s exposures in a 1 minute cadence, at $g \sim 23$ magnitude limits, a wide-field ($\sim$ 3 deg$^2$) of view (a comparison of DWF with other surveys can be found in Table \ref{survey_table}).  We note that the 1 minute cadence and 20s exposure times are due to overheads in inefficient readout times.
In conjunction with optical observations carried out with DECam on the 4m Blanco Telescope in Chile, wide-field ground- and space-based observatories spanning the entire electromagnetic spectrum are coordinated either to simultaneously collect data on the same region of the sky, or coordinated to trigger rapid (or later-time) follow-up of transient sources.  The DWF program is carried out for one week twice annually.
Data collected by DECam for real-time analysis is highly compressed \citep{dwf_data_comp}, to minimize transfer speed, and sent directly from the summit on Cerro Tololo, Chile to the OzSTAR supercomputer at Swinburne University of Technology in Australia for processing and analysis. In addition, these data are also transferred using lossless compression and fully processed by a modified version of the photpipe NOAO Community Pipeline \citep{noao_photpipe,noao1,noao2} at a later time.
The DWF program, like many other transient surveys \citep[e.g., PTF and SN Legacy Survey, among others,][]{ptf_imsub,snlegacy_imsub}, relies on an image subtraction pipeline \citep[\textit{Mary} pipeline,][]{mary} to detect potential sources of interest in real time.  A ranked list of candidates is presented to astronomers and volunteers for further visual inspection of image cutouts (small fraction of the DECam FoV centered on a single detected source) and light curves using the interactive tools described in \cite{dwf_visualization}.

\subsection{DWF Data Samples} \label{sec:data}
Here, we describe the DWF data stream in more detail, the light curve creation process, and the final inputs that will be fed into the transient identification algorithm.  The data collected by DECam for the DWF program is unique among transient surveys in its cadence, and therefore offers the potential for "first of their kind" discoveries.  For DWF, the 4-meter Blanco telescope, on which DECam is mounted, collects continuous 20s exposures at $\sim$ 1 minute cadence, when including readout time.  In each 20s exposure, DECam reaches a depth of $g \sim 23$ under normal DWF observing conditions, $\sim 1.0$ arcsecond seeing at $\sim1.5$ airmass. The slightly higher than ideal airmass is due to the visibility requirements for simultaneous observations in the radio, conducted by telescopes in either Australia or South Africa, and telescopes operating at other wavelengths in the Antarctic, North America, and other locations, including space-based telescopes.

The $g$-band is selected as the main observing band for DWF as DECam sensitivity is $\sim 0.5$ magnitudes deeper than in redder filters, many fast bursts are hot and blue, and DWF target fields are typically at low Galactic extinction. 
Most DWF target fields have template reference images taken prior to the run in multiple filters. In addition, and if there are no reference images (i.e., for newly discovered FRB or short-GRB fields), the target fields are observed either at the start or end of the night (or both) in other filters, typically $r$- and $i$-bands to determine source colors.

The DWF program collects data with DECam over a $\sim 3$ deg$^2$ FoV.  This wide-field is covered by the 62 individual DECam CCDs.  The data from each CCD is saved as an extension in a multi-extension fits file.  These data are processed and analyzed in two ways. 

Firstly, for real-time, or fast analysis, the image files are `lossy', compressed at the summit using the method described in \cite{dwf_data_comp}, and sent to  OzSTAR supercomputer at Swinburne University of Technology for data analysis. Data transfer from the Cerro Tololo summit in Chile to Australia is too slow to enable data processing, analysis, and transient candidate identification within minutes, which is necessary for fast transients. The lossy compression is tunable to the speed of the internet and can speed up transfer by compressing the data up to $\sim$20$\times$ and still enable detection of $\sim$ 95\% of the transients.  Furthermore, to enable fast identification and rapid response follow-up triggers, these data are `fast' processed in parallel on the OzSTAR supercomputer.  The fast processing sacrifices some aspects of a full processing pipeline for speed.  Both the lossy compression and the fast processing result in several artifacts in the images that are not typically observed in conventional transient pipeline analyses.

The real-time data processing for the data collected on the dates used in this work includes using Swarp \citep{swarp} to align and stack images, SExtractor \citep{sextractor} to identify sources, and HOTPANTS \citep{hotpants} to perform image subtractions.  After performing image subtraction and source extraction on the differenced images, the Mary pipeline \citep{mary} runs a machine learning algorithm on the potential candidates to minimize CCD artifacts.  Aperture photometry of one full-width half-max was forced on the coordinates of sources that contained a residual following an image subtraction.  The remaining candidates are then ranked based on their presence in The Second-Generation Guide Star Catalog \citep[GSC-II,][]{gsc2} and in previous nights of the DWF run, with higher rankings given to those sources that are present in neither GSC-II, nor previous DWF nights.  Data analyzed in this manner will be referred to as ``real-time'' data.  We note that the real-time processing is different for later runs.

Secondly, the data are separately sent to the NOAO High-Performance Pipeline System \citep{noao1,noao2} to provide post-run, fully processed and well-calibrated data for later-time analyses.  These data are used for fast transient detection after burst, fast transient searches cross-matched with other wavelengths, fast transients associated with slower-evolving events (e.g., supernova shock breakouts), slower-evolving events caught early by DWF, and other applications.  For the data used here, sources were identified using SExtractor 
and the images are not stacked, nor image subtracted, however.  Automatically calculated apertures were forced on the coordinates of all sources $1.5\sigma$ greater than the background.  Magnitudes from SExtractor identified sources are calibrated against the SkyMapper Data Release 2 catalogue \citep{skymapper}. Data analyzed in this manner will be referred to as ``post-run'' data. 

For both real-time and post-run data processing methods, the light curves are generated for 
sources that have one or more detections at the same coordinates using aperture photometry on non-subtracted images; DWF targets are named using these coordinates.  For each DWF source, a data point or upper limit is generated every $\sim$ 1 minute, unless the source location falls off the CCD, either in chip gaps or off the edge of DECam FoV as a result of small offsets in guiding, tracking, and hexapod tip-tile corrections, as a result of changing weather, moving to a new field, etc.

\begin{deluxetable*}{ccccc}
\tablenum{2}
\tablecaption{DWF Runs Analyzed with FTF Algorithm\label{datatable}}
\tablewidth{0pt}
\tablehead{
\colhead{Field Name} & \colhead{RA (center)} & \colhead{Dec (center)} & \colhead{Start Date} & {End Date}
}
\startdata
\multirow{2}{*}{4-hour} & \multirow{2}{*}{04:10:00}&\multirow{2}{*}{-55:00:00} & 2015-01-14  & 2015-01-17\\&&&2015-12-18& 2015-12-22 \\
\multirow{2}{*}{Antlia$^\dagger$}&\multirow{2}{*}{10:30:04}& \multirow{2}{*}{-35:19:24} & 2018-06-07 & 2018-06-09 \\&&&2017-02-03&2017-02-07\\
FRB010724 & 01:18:00 & -75:12:00 & 2015-12-18 & 2015-12-22 \\
CDFS Legacy$^+$& 03:30:00&-28:06:00&2019-12-03&2019-12-07\\
FRB171019$^+$& 22:17:32&-08:39:32 & 2019-12-05&2019-12-07
\enddata
\tablecomments{DWF fields analyzed as a part of this study.  Fields noted with a $+$ are those with real-time data.  Fields noted with a $\dagger$ are those that have been analyzed in \cite{dwf_ul}.}
\end{deluxetable*}

There are a total of 5 DWF fields analyzed in this work, shown in Table \ref{datatable}.  There are 2 ``real-time'' data sets covering the CDFS Legacy and FRB171019 fields.  There are 5 ``post-run'' data sets covering two epochs on the 4-hour and Antlia fields and one epoch on the FRB010724 field.  The two 4-hour and Antlia epochs analyzed here are from two separate runs, spaced 11 months and 16 months apart, respectively; this second pointing can help establish if there is any recurrence or periodicity to transient behavior observed.  

The 4-hour field is one of the first fields observed by DWF.  The first DWF run employed an observational routine with dithering.  Analyzing the first run on the 4-hour field can determine how robust our algorithm is to dithered data; subsequent DWF runs have moved away from the dithered approach due to confounding issues discussed in Section \ref{realtimeresults}.  The Antlia field was chosen for analysis, in part because comparisons can be drawn between this work, and work done in \cite{dwf_ul}.  The FRB010724 data is from a dense field with 839,729 light curves generated over 5 days.  The two ``real-time'' fields were chosen out of necessity; older ``real-time'' data was not stored for later analysis, and the COVID-19 pandemic, which has halted operations for many observing sites across the world, had precluded the acquisition of ``real-time'' data sets from Cerro Tololo for several months.  The results of running the Fast Transient Finding (FTF) algorithm on the 5 data sets in Table \ref{datatable} are presented in Section \ref{results}.  The naming convention for the light curves presented in this paper are the survey name, DWF, followed by the right ascension (RA) and declination (DEC) in sexagesimal coordinates as follows: \textbf{DWFRADEC}.

\subsection{Challenges of Detecting and Studying Fast Transients}

The challenges of ``big-data'' in astronomy have been well documented \citep[e.g.,][]{bigdata1,bigdata2,bigdata3,bigdata4}. As seen in Table \ref{survey_table}, the cadence of many optical transient surveys allows for longer processing times, but could limit the speed with which astronomers detect transient phenomena, with potential delays of several days between the start of an event and its detection.  DWF offers a different approach to other optical surveys, and presents new challenges to analyze incoming data in ``real-time''.  The real-time, fast processing by DWF (described in Section~\ref{sec:data}) is by no means ideal or optimal;  the lossy compression adds artifacts and the fast data processing is much poorer than normal processing, creating additional artifacts and poorer astrometry and alignments, which can yield poorer subtractions.  This sub-optimal fast processing is necessary, however, in order to identify events and trigger follow-up before sources fade.  Detailed follow-up of these events, ideally spectra, can shed light on the early phases of SNe, GRB afterglows, potential FRB optical counterparts, and other transient phenomena.  The challenges outlined here are unique to DWF due to the fast cadence and the opportunity for transient detection on minute time-scales.  It is these challenges that motivate the work presented here.

\section{Algorithm for Early Source Detection} \label{algo}

Despite its ubiquity, the use of image subtraction techniques to identify transient
sources is wrought with challenges.  The convolution of point spread functions between images can
be challenging, if not impossible between different instruments and different
seeing conditions; even when feasible, convolution can be computationally intensive. 
Source extraction codes (or the astronomers interpreting their outputs) can be fooled by sources that are not real, for example cosmic rays, cross-talk on images, or mis-aligned subtractions.  As surveys search for transients to fainter magnitudes, they begin to hit the noise/source threshold and many detections are too ambiguous to accurately identify.
The very large number of source detections in image subtracted frames and the inability to have humans analyze them all in a reasonable time frame (especially for fast transients), or do so with any solid accuracy to trigger followup observations, necessitates the use of machine learning frameworks to identify false-positives
\citep{ipac_imsub,toros_imsub,ztf_imsub}, further complicating the process and
increasing computational demands. 
 
Furthermore, the machine-learning (ML) approach requires extensive 
training and large training set samples (typically in the hundreds of thousands of images), increasing the demands on human time and capital.  The challenges
associated with image subtraction 
have led to attempts to identify transient
sources through direct image
comparisons \citep{imsub_alt1} and light curve analysis 
\citep[e.g.,][]{gaia_fast,imsub_alt2,dwf_ul}.

In addition, definitive source classification is hardly possible with image subtraction alone:  transient characterization is confirmed through follow-up observations, including spectroscopic data; however, telescope time using sensitive spectroscopic instruments is very limited, and the time-window for observing fast-fading sources is narrower than the cadence of conventional transient surveys.  If enough data of the source is rapidly available for a light curve to be made (e.g. within a few minutes
of the first data acquisition), preliminary classification can be performed, using simple metrics such as its rise and/or decay rate and peak brightness.  This early classification using light curves can inform astronomers about the resources they should allocate to these targets which are always limited; this kind of observation strategy will be crucial for LSST, and is the service that brokers will be performing for the community \citep[][]{lsst_alerce,lsst_mars,lsst_antares,lsst_lasair,lsst_ztf,lsst_fink}.  
With better sampling, including light curves spanning multiple wavelength bands, a more precise classification of sources can be achieved \citep{mlclass1,mlclass2,mlclass3,mlclass4,mlclass5,mlclass6}, but progress on this front is still minimal due to the complexity of the data and classification algorithms.

\subsection{Fast Transient Finding (FTF) Algorithm} \label{sec:FTF}
\begin{figure}[h]
    \centering
    \plotone{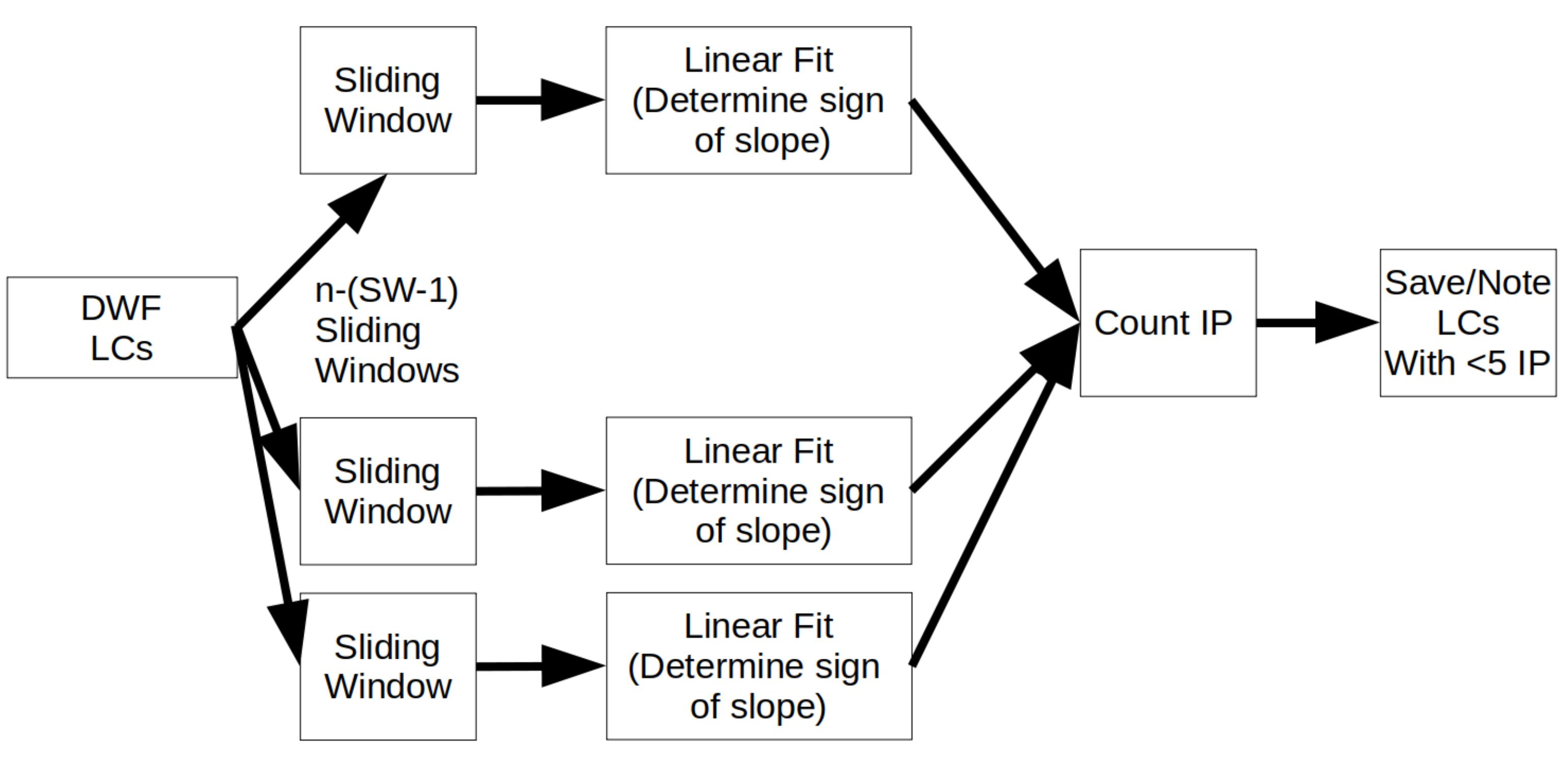}
    \caption{For a given DWF field, a total number of $N$ sources are detected.  A light curve (LC) will be generated for each source during post-run processing, or in real-time for those sources identified as potential transients from image subtraction (see Section \ref{sec:data} for a thorough discussion of the different data types). Each LC is fed into the algorithm described in Section \ref{sec:FTF} in a parallel-processed manner.  The LCs are separated into $n-(\rm{SW}-1)$ different sliding windows, where $n$ is the number of data points in the LC and SW is the size of the sliding window.  Each of these sliding windows are processed in parallel, fit linearly, and the sign of the slope is determined: positive, negative, or flat.  The signs from the slopes of the individual sliding windows is recombined and the number of inflection points (IPs) in the LC is counted.  Those LCs with fewer than 5 IPs are saved as potential transients.}
    \label{algo_flowchart}
\end{figure}

Given the obstacles inherent in using image subtraction techniques, and the necessity of a light curve analysis to classify peculiar transient events, we propose to identify these transient phenomena with a direct light curve analysis of the DWF data stream.  The algorithm we describe can  be used as an independent verification for candidates detected via other methods (e.g. image subtraction, machine learning algorithms);
a flow chart of the FTF algorithm is shown in Figure \ref{algo_flowchart}.

For each unique source observed during the DWF run, we separate the light curve data using a sliding window (SW), a technique common in financial time series analysis \citep{financial_sw1,financial_sw2,financial_sw3} as well as machine learning applications across several disciplines \citep{sw_ml1,sw_ml2,sw_ml3,sw_ml4}.  The user can define the size of the sliding window parameter, but is limited by the number of data points contained within an individual light curve file (light curves may have missing points due to changing weather conditions, upper limits, or artifacts that prevent our photometric pipeline to accurately estimate the magnitude of the source).   The source code for the FTF algorithm is publicly available\footnote{https://github.com/rstrausb/FTF}.

\subsection{Statistical Selection of Algorithm Parameters} \label{stat}

Based on the typical field cadence and the number of points per light curve we can assess the best SW size. We emphasize here that, while we focus in this paper on finding known categories of fast evolving transients (e.g., GRB afterglows, kilonovae, etc.) the FTF algorithm can be easily customized for different or novel types of variable phenomena by changing the sliding window size (Figure \ref{counthistograms}) and the slope threshold (Figure \ref{slopehistograms}); searches for new types of fast-evolving transients is an important focus of DWF and we will us the FTF to pursue these targets in the future.

In Figure \ref{counthistograms}, we present the histograms of the number of data points present in each light curve for all of the fields and runs analyzed in this work, for both the ``real-time'' and ``post-run'' data.  The data points for both the real-time and post-run are generated by forced photometry.  In the real-time data, aperture photometry of one full-width half-max was forced on the coordinates of sources that contained a residual following an image subtraction.  In the post-run data, automatically calculated apertures were forced on the coordinates of all sources $1.5\sigma$ greater than the background. The red dashed line in Figure \ref{counthistograms} represents our choice of a SW=5; this avoids the predominance of noisy light curves (${\rm SW} \ll5$) and mitigates the risk of averaging over rapidly rising and falling light curves or flares with larger windows (${\rm SW} \gg 5$).

\begin{figure}
    \centering
    \plottwo{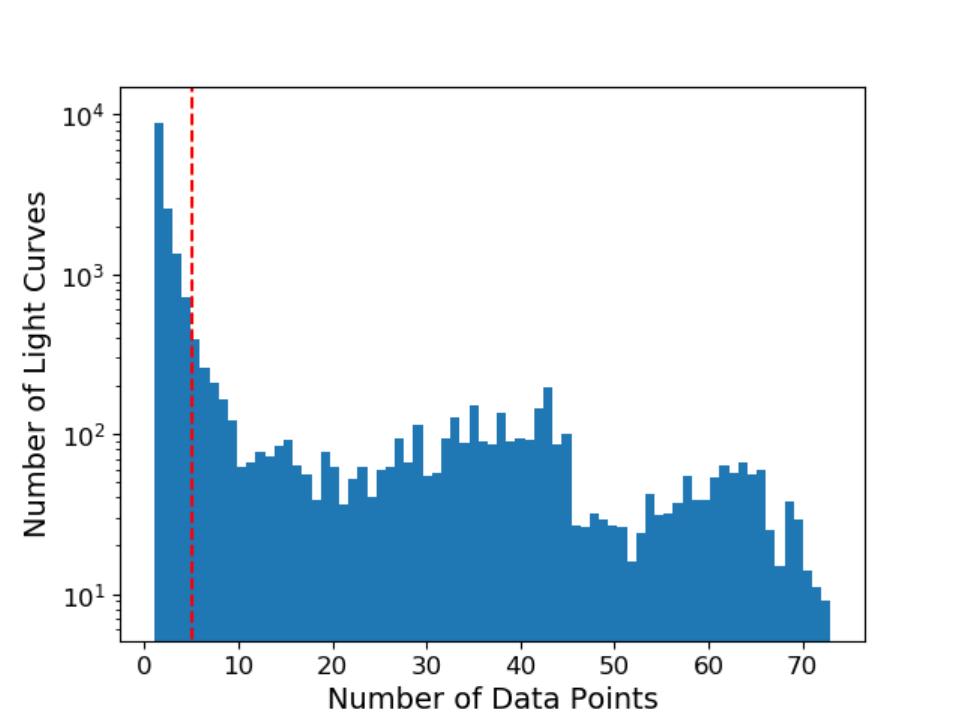}{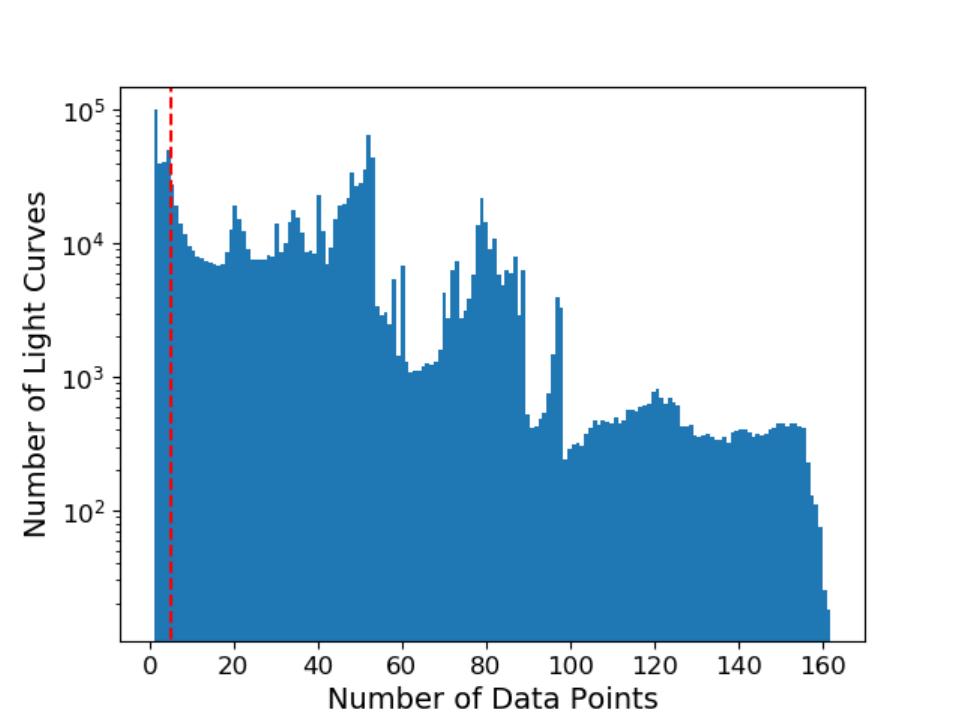}
    \caption{Amalgamated histogram of the number of data points present in each light curve for all of the fields and runs analyzed in this work in both the real-time data (left) and post-run processed data (right) sets.  The red dashed line is plotted at 5, the chosen sliding window size in the FTF algorithm explored in this paper; the sliding window parameter can be changed to search for transients evolving on different time scales.}
    \label{counthistograms}
\end{figure}



Over each sliding window, we compute a simple linear fit, $g = \alpha t + g_0$ (where $g$ is the observed magnitude, $t$ is the time in minutes from the first observation, $\alpha$ is the temporal decay  index, and $g_0$ is the intercept), and return the slope and its uncertainty.  We do not perform our fit using uncertainties in the photometry for reasons of efficiency.

A histogram of the slopes of each window from real-time and post-processed fields is plotted in the two graphs in Figure \ref{slopehistograms}.  We find that the distributions of slopes is well-modeled with a Laplace Distribution, represented by the probability density function 
\begin{equation}
P(x)=\frac{1}{2b} \exp{\left(\frac{|x-\mu|}{b}\right)}
\label{laplaceeqn}
\end{equation}

where  $\mu$ is the mean (which, in the case of this function is equal to the median as well as the mode), the variance is $2b^2$, and the average absolute deviation is $b$.  From the linear fits, we obtain the sign of each window slope as positive, negative, or flat; we consider a flat (non-changing) slope if $\alpha=0 \pm b$, 
as shown by the red dashed lines in Figure \ref{slopehistograms}. 
The algorithm keeps track of the sign of the slope over each sliding window and notes a change in the sign of the slope as an inflection point (IP).  Scanning over each sliding window, the algorithm tallies and records the number of IPs.
For example, a typical fast previously unknown transient may have a number of IPs between 0 (straight rising or decay behavior) and 3 (e.g. a flare with one IP rising, one IP fading, and one IP flat).

\begin{figure}
    \centering
    \plottwo{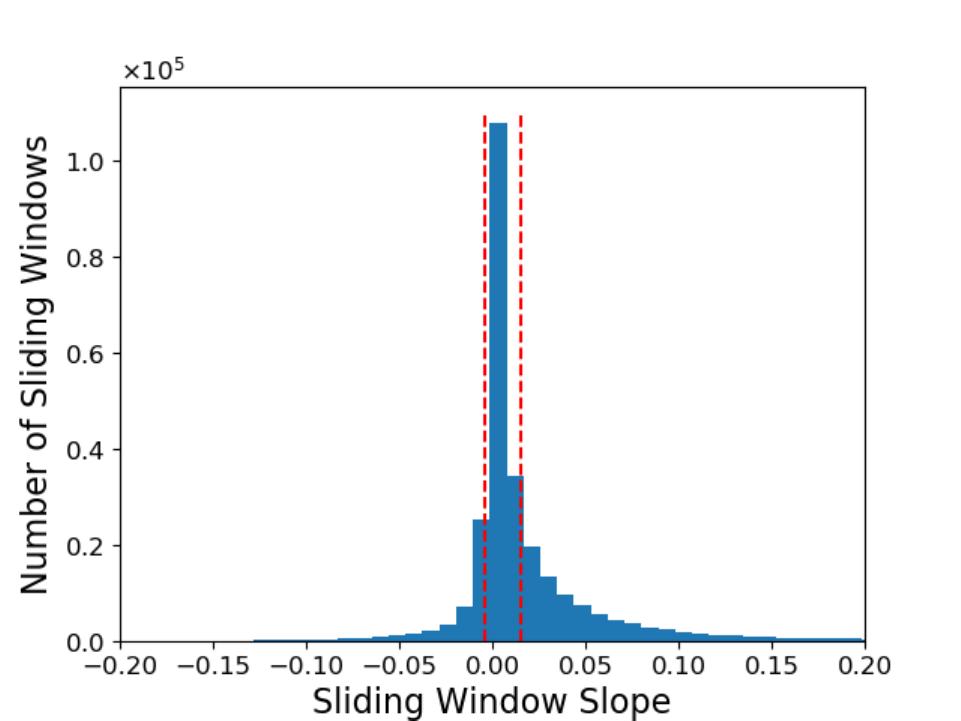}{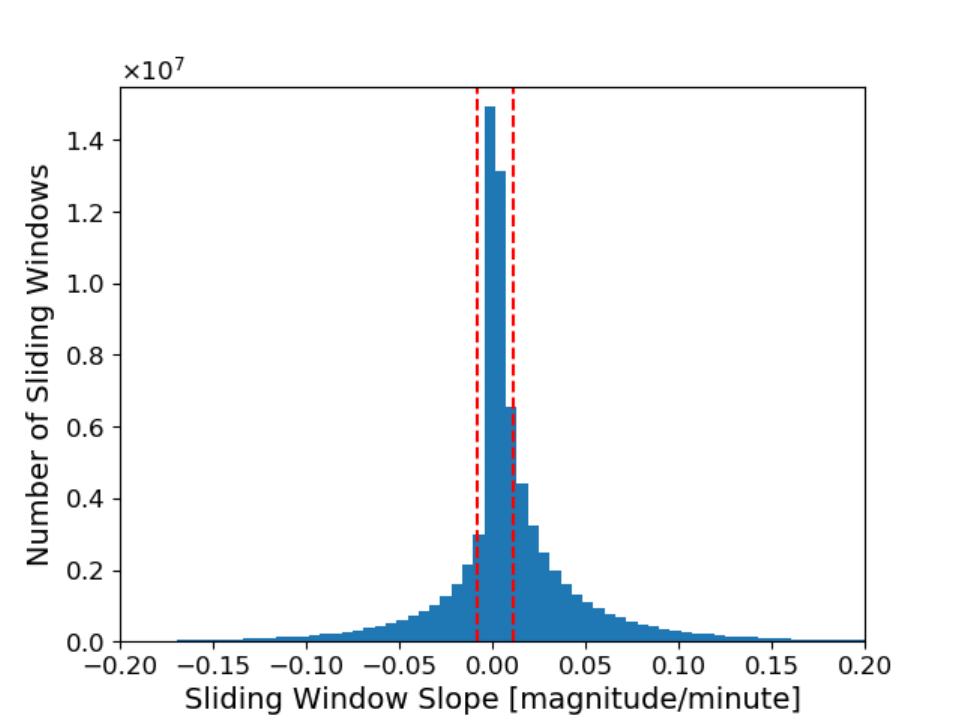}
    \caption{Histograms of the slope of each separate sliding window for the real-time data (left) and the post-run processed data (right).  We fit the histogram with a Laplacian Distribution, as defined in Equation \ref{laplaceeqn} and plot the $b$ term, with a red dashed line; the slopes within the red dashed lines are considered ``flat'' in the FTF algorithm explored in this paper; this slope threshold can be tuned to search for transients with different intensities, as shown by the blue dashed line in Figure \ref{flare}.}
    \label{slopehistograms}
\end{figure}


The aforementioned process is time consuming and CPU-intensive. Since our ultimate goal is to provide real-time identification of fast transients from the DWF data stream we implement a full parallelization of our python-based algorithm; each target can be run independently and in parallel during the DWF run.  This parallelization enables the code to run over each DWF source in $\lesssim 1$ minute, on par with the cadence of incoming data points.  The efficiency of the code is important for real-time identification of transients, especially deployed on multi-core CPUs, like the supercomputer at Swinburne University, where the optical data from DWF runs is analyzed.  

\subsection{Phenomenological Selection of Number of Inflection Points} \label{sec:FTFIP}

For each light curve we calculate the number of IPs and then we group all objects and light curves with the same number of IPs. For this work, we focus on light curves that have four or fewer IPs within the typical DWF light curve ($\sim$1 hour).

\begin{itemize}
    \item Light curves with zero IPs, but that are monotonically increasing or decreasing could be longer evolving transients: Cepheids, RR Lyrae, or SNe, for example.  These could also be the beginning a of fast evolving transient at the end of DWF observation.
    \item Light curves with one IP might be catching the start of the rise or fall of a transient evolving on minutes to days time scales.
    \item Light curves with two or three IPs may contain peaks or dips spanning the entire DWF time on the field (typically 1-4 hours). 
    \item   Four IP light curves could point towards more complex behavior that goes through several phases over the course of the DWF observation.
\end{itemize}

It is important to note that transient phenomena may be occurring before the DWF run began, or continue after data acquisition has stopped.  Therefore, a burst like event might only have two or three IPs, as its light curve might be shifted towards the beginning or end of a run in such a way that some parts of the curve are not sampled.

For the FTF algorithm, we define ``fast transient'' candidates as those with fewer than four inflections points, and with at least one sliding window with a slope greater than a user specified slope; in this iteration of the algorithm, that specific slope is defined by statistical measures as defined in Section \ref{stat}, but could be set manually by the user if searching for a specific type of source with a known range of slopes.  The code could be modified to include sources with more than four inflection points points towards a variable source that could be of interest to other areas of astronomy.

Using this algorithm, the first potential transients can be reported just after the first five minutes of observation by DWF; thereafter the number of IPs associated with each light curve will be updated every minute, as the sliding window shifts over by one data point.  A noted inflection change with a corresponding positive detection in the image subtraction pipeline provides good evidence to trigger imaging and spectroscopic follow-up.


\subsection{FTF Algorithm Demonstration} \label{demo}

\begin{figure}[h]
    \centering
    \plottwo{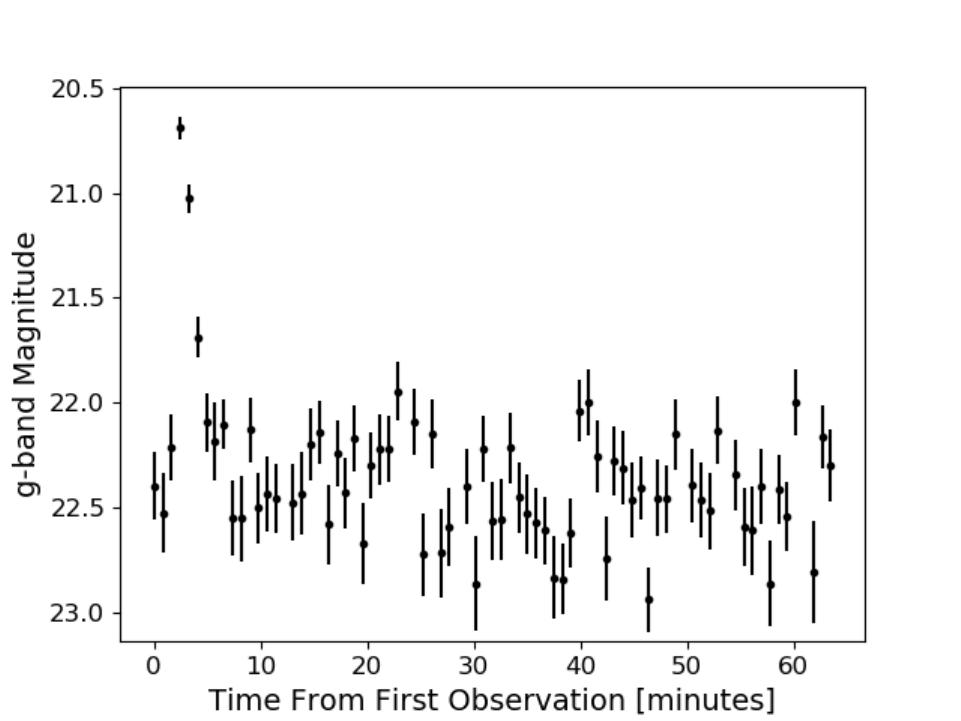}{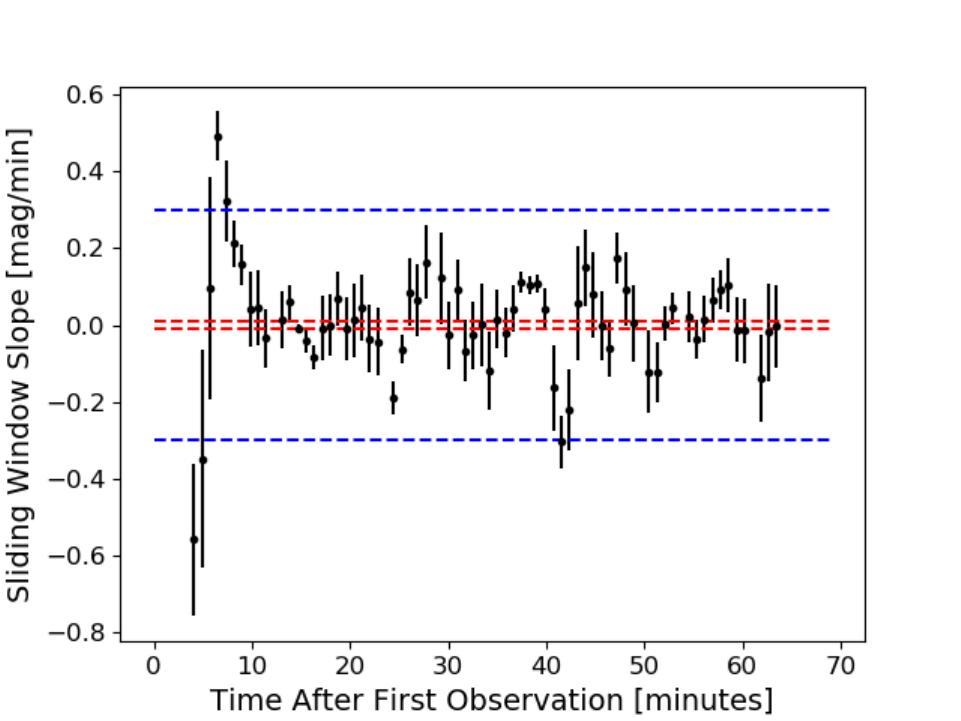}
    \caption{Left: Light curve of a flaring star, DWF102955.559-360035.170, observed in the Antlia field by DWF.  This object was first discovered using unsupervised learning techniques in \cite{dwf_ul}.  Right: The slope of each sliding window is plotted over time.  The red dashed line correspond to the $b$-term derived from the Laplace Distribution, as defined in Equation \ref{laplaceeqn} and plotted in Figure \ref{slopehistograms}.  The blue dashed line represents a more restrictive parameter that can be tuned in the algorithm if for example, astronomers are looking specifically for flare stars.}
    \label{flare}
\end{figure}

In Figure \ref{flare} we show how the FTF algorithm works on a sample light curve, using a flaring star first detected in \cite{dwf_ul} as an example.  The light curve for this flaring source, DWF102955.559-360035.170, is plotted in the left panel of Figure \ref{flare}.
The right panel in Figure \ref{flare} shows the slope derived from each sliding window as function of time. We can clearly see that the flare in the light curve and the relative inflection points enable the identification of a change in brightness beyond the typical brightness. While this information may be used to trigger follow up observations, the subsequent data demonstrate that there are more inflection points and therefore the source is not a fast transient as classified in Section \ref{sec:FTFIP}. 
The dashed red lines in the figure represent the same thresholds identified in the histograms presented in Figure \ref{slopehistograms}; users can set a different threshold to identify different transients of interest, as shown by the blue dashed lines.

\section{Results} \label{results}

In this section we present the outcome of our FTF algorithm and the implication on 1) detectability of fast transients of different natures and the rate of detection of these objects compared to other surveys and 2) the required effort for spectroscopic follow-up for secure classification. In summary:

\begin{itemize}
    \item On a single night of observation on a single field we obtained on average $\sim 50,000$ real-time light curves, and $\sim 340,000$ post-processed light curves. 
    \item Feeding these $50,000$ real-time and $340,000$ post-processed light curves into the FTF algorithm, we detect, on average $150$ ($\sim 3\%$) and $3,000$ ($\lesssim 1\%$) potential fast transients respectively.
    \item Checking the science frames of the potential fast transients for artifacts and other non-astronomical sources, we obtained on average, 1 statistically significant fast transient per field in the real-time data, and 13 statistically significant fast transients per field in the post-processed data.
    \item Based on light curve fits 69 sources can be classified as ``potential transients'' under our definition in Section \ref{sec:FTF} from the fields described in Table \ref{datatable}.
    \item In the event of a fast transient identification, FTF would allow a latency time of just 5-15 minutes for multi-wavelength and spectroscopic follow-up observations.
 \end{itemize}

A detailed break-down of the results over each field can be found in Table \ref{results_table}.  We note that the 69 sources identified as ``potential transients'' are fast evolving sources that would require follow-up, specifically spectroscopic follow-up to determine if these sources are indeed transients or other variable sources.
%

\subsection{Results on Real-Time and Post-Run Data} \label{realtimeresults}

Once a list of candidates is generated using the FTF algorithm (within the first 5-10 minutes of the run, and then every minute thereafter), as shown in Section \ref{demo}, light curves are vetted by our team; for those light curves that passed human inspection, image cutouts for the source's location on the sky were visually inspected to exclude the presence of artifacts that survived our processing pipeline (e.g., cosmic rays, bad pixels, bad rows/columns, etc.).  For the purposes of this paper, a positive detection is defined as one where the source is either a known variable, 
a DWF variable detected by other methods \citep[see for example,][]{dwf_ul}, or a newly discovered candidate that passes a visual inspection of the images associated with the light curves.  To confirm known variables, we checked the coordinates of our candidates against known variable source catalogs such as the General Catalog of Variable Stars \citep[][]{gcvs} and the International Variable Star Index \citep{vsi}.

\begin{deluxetable*}{lrrr}
\tablenum{3}
\tablecaption{FTF Results\label{results_table}}
\tablewidth{0pt}
\tablehead{
&\colhead{Number of}\vspace{-0.4cm}&\colhead{Algorithm Identified}&\colhead{Human Filtered}\\\colhead{Field Name}&\vspace{-0.4cm}&&\\
& \colhead{Light Curves} & \colhead{Light Curves} & \colhead{Final Results}
}
\startdata
4-hour (Run 1) & 155,989 & 758  & 20 \\4-hour (Run 2)& 182,486 & 448 & 10 \\
Antlia (Run 1)&350,542& 2,001& 7 \\Antlia (Run 2)&161,511& 1,364 & 7 \\
FRB010724 & 839,729 & 5,512 & 23 \\
CDFS Legacy*& 84,394 & 1,37& 1\\
FRB171019*& 17,209 & 265 & 1 \\
Total & 1,791,860 & 15,897 & 69
\enddata
\tablecomments{The FTF algorithm identified $\approx 1\%$ of the light curves in the fields studied as potential transients, reducing the number of real-time light curves that require human inspection by two orders of magnitude.  From those light curves identified by the algorithm, about $0.5\%$ are identified by a human observer as potentially real astrophysical phenomena, after rejecting sources with obvious non-astrophysical explanations. Fields with an * denote fields where real-time data was analyzed.}
\end{deluxetable*}





\begin{figure}[h]
    \centering
    \plottwo{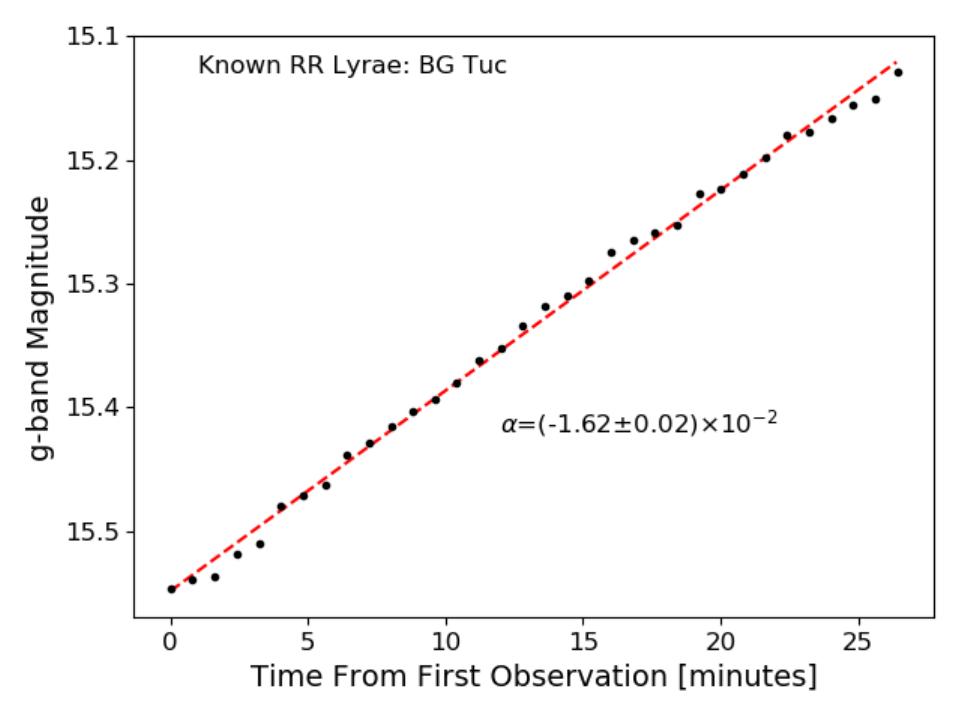}{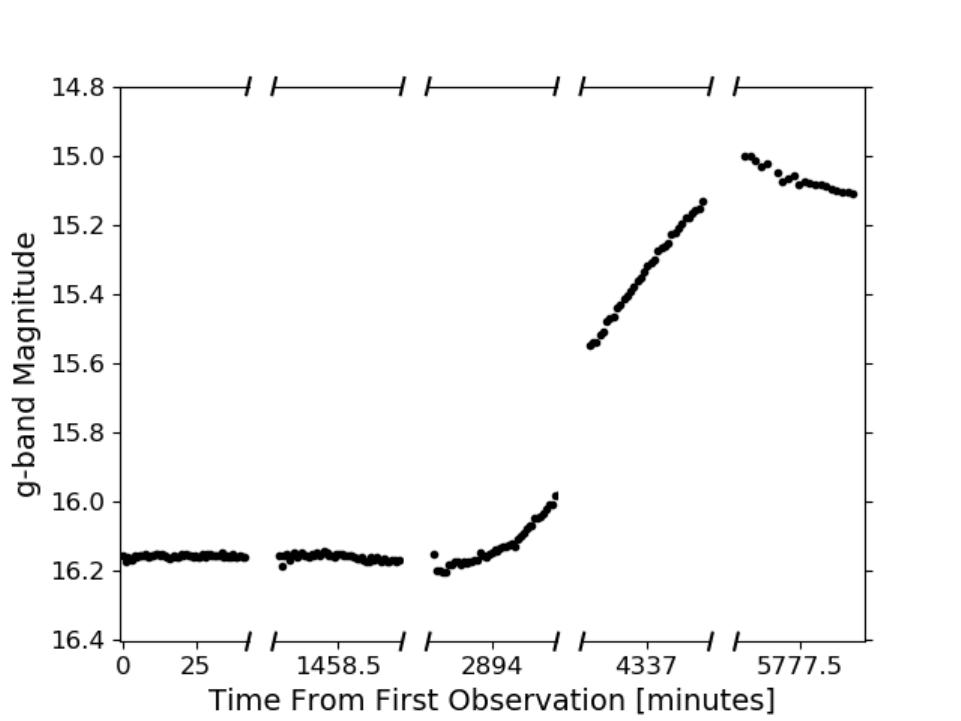}
    \caption{Left: Light curve of DWF011805.113-751125.458 that was detected by the FTF algorithm; its slope labeled, $\alpha$, is plotted in the red dashed line.  These coordinates correspond to the known RR Lyrae, BG Tuc.  Right: Data from the other DWF nights are plotted for DWF011805.113-751125.458, showing the variability of this object over the week-long DWF run.  The section plotted in the left panel corresponds to the second to last section of the right hand graph.}
    \label{bg_tuc}
\end{figure}

\begin{figure*}[h]
 \gridline{\fig{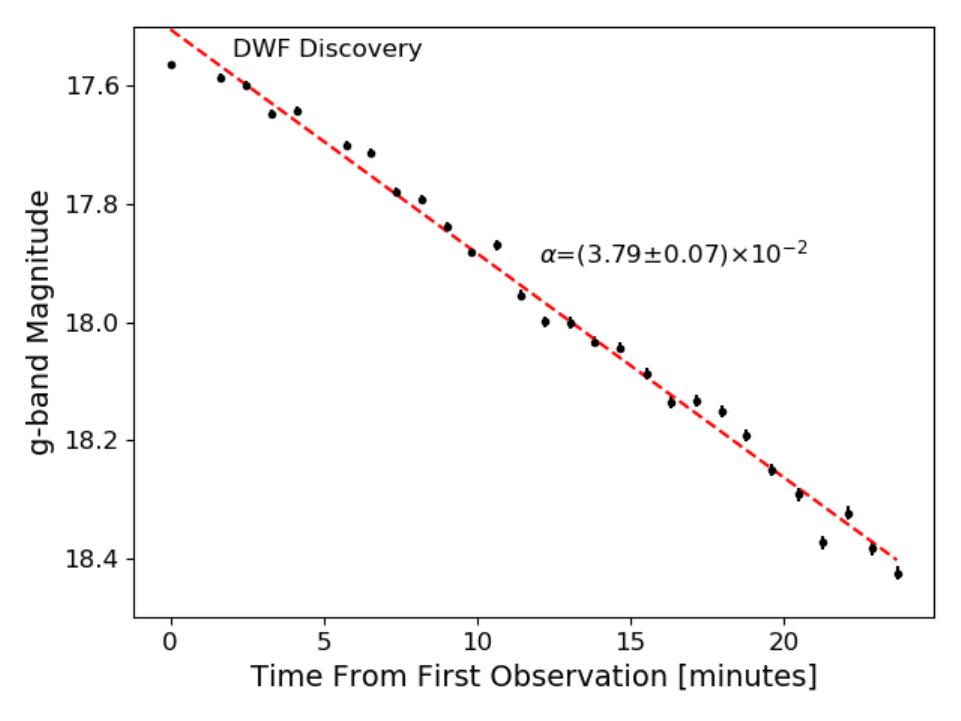}{0.45\textwidth}{}
          \fig{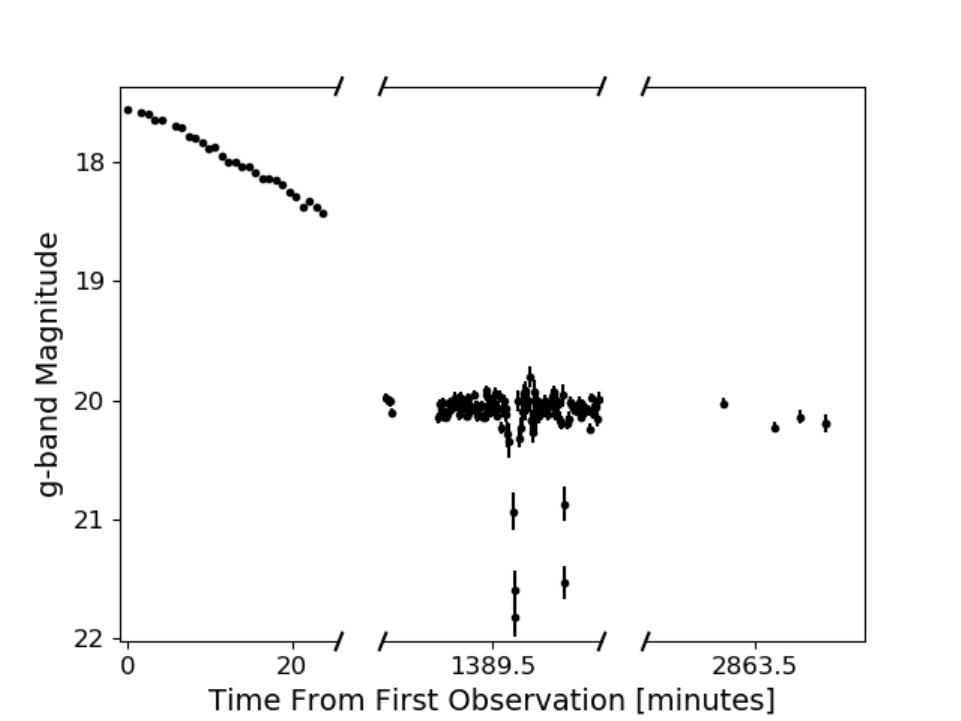}{0.45\textwidth}{}}
 \gridline{\fig{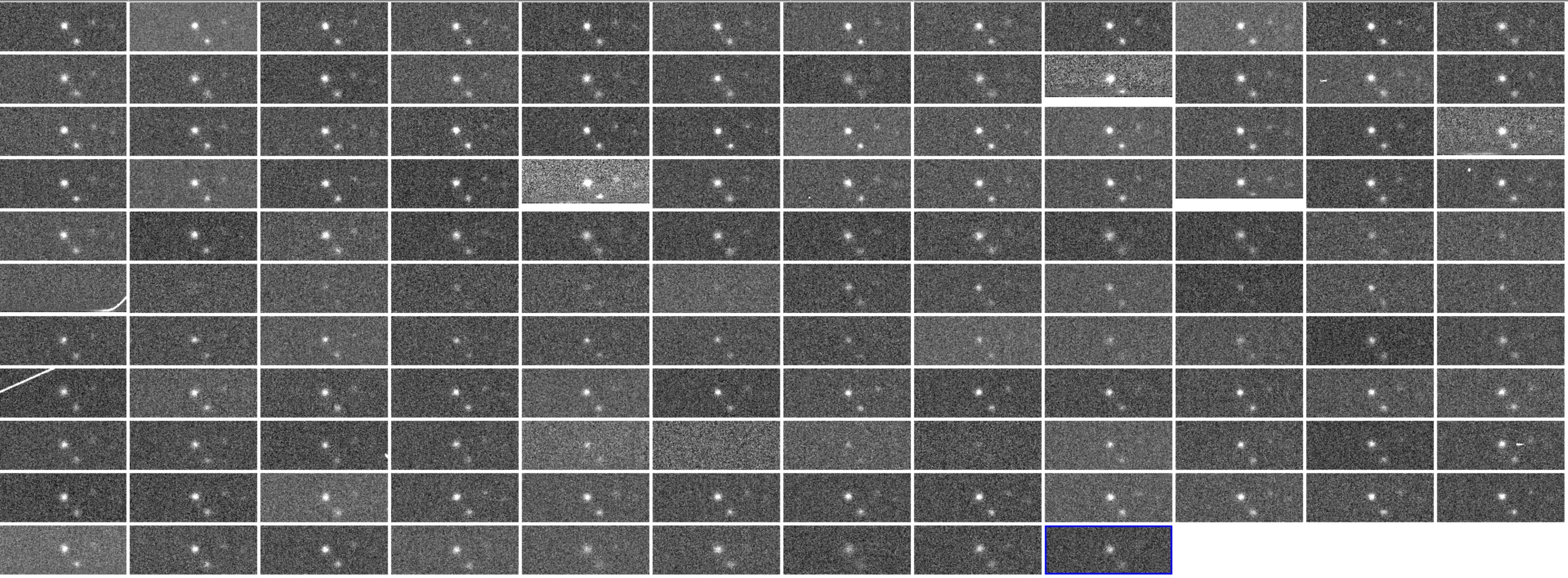}{0.75\textwidth}{}
 \label{doubledip}
          }
          
 \caption{Top Left: The light curve of DWF102920.187-355700.211 is plotted for the night of 180607.  There does not appear to be contamination from non-astrophysical effects in the image cutouts from this night of data, so we identify this source as real.  Top Right: The light curves for the other nights for which this source is observed.  Overall, the source is declining form $g\sim17.5$ in the first night before down to an almost constant $g\sim20$ magnitude in the second and third night.  There are however dips of about 2 magnitudes present during the second night.  Bottom:  The image cutouts for the second night of data from this source are presented.  As can be seen in one of the middle rows, the source and those nearby all seem to fade, indicative of clouds that may not have been visible to the astronomers on the ground.}
 \label{fading}
 \end{figure*}

\begin{figure}
    \centering
    \plottwo{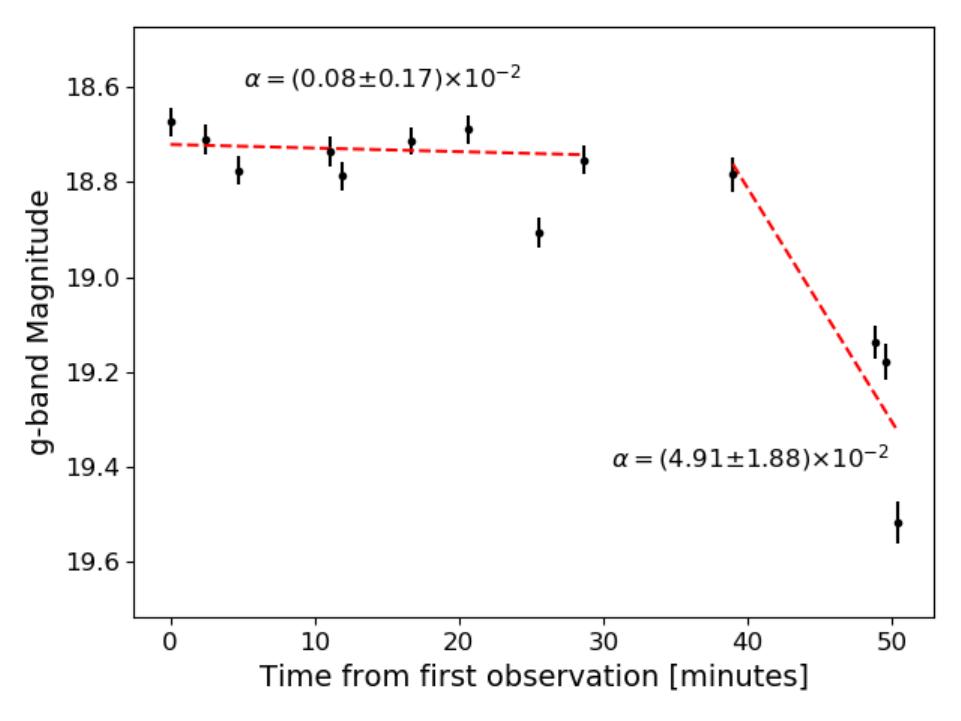}{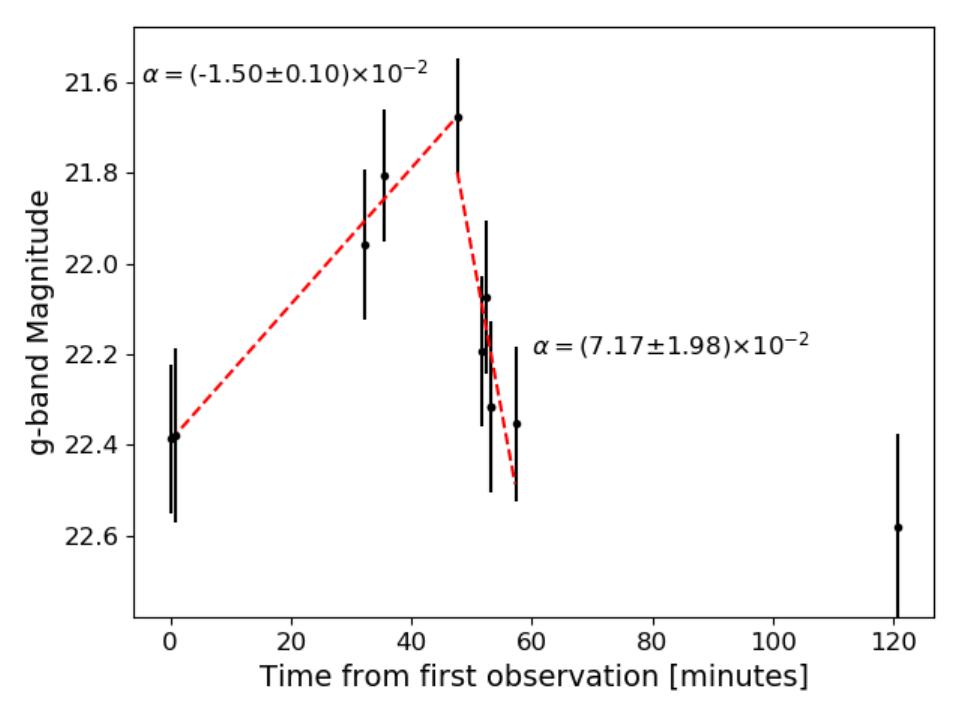}
    \caption{Left: The light curve of DWF040623.456-550041.171 is presented.  The first 10 data points in the light curve are consistent with the flat slope defined in Section \ref{sec:FTF}.  The source then dims by 0.8 mangitudes in about 10 minutes.  After the first data point in this decay, the FTF algorithm will note this object as a potential fast transient. Right: The light curve of DWF102613.233-350150.332 is presented.  The first 5 data points in this light curve would indicate this object as a potential fast transient.  The light curve then enters into a period of seemingly exponential decay for the remainder of observation.  The algorithm would detect this inflection change, again alerting astronomers about the potential transient nature of this source.  In both cases present in this figure, the FTF algorithm can alert astronomers within 5-10 minutes of the transient behavior.}
    \label{inflex}
\end{figure}

It is important to note that the real-time light curves will only exist for those sources that were candidates identified via image subtraction as a part of the Mary pipeline analysis; in contrast, the light curves from the post-run processing with the NOAO pipeline encompasses all sources that were detected during the run.  The sources in both real-time and post-run data sets include point sources and extended sources.  The linear fits plotted in the subsequent figures (Figures \ref{bg_tuc}-\ref{edge2}) are meant to give an idea of the general trends in the lights curves, and are not the slopes associated with the sliding windows, as shown in Section \ref{demo}, nor are they necessarily the best fit for the data.

The light curve of DWF011805.113-751125.458, plotted in the left panel of Figure \ref{bg_tuc} is a known RR Lyrae source, called BG Tuc \citep{bgtuc1,bgtuc2}.  The light curves of DWF011805.113-751125.458 for each night of the DWF observing run are plotted in the right panel of Figure \ref{bg_tuc}; this figure shows the variability of the object over long time scales.

The behavior of DWF102920.187-355700.211 on both the night of 180607, plotted in the upper left and panel of Figure \ref{fading}, and 1800608, plotted in the upper right panel of Figure \ref{fading}, 
were identified by the FTF algorithm as potential transient phenomena.  The first night of data shows a source decreasing from $g\sim17.6$ to $g\sim18.4$ in 30 minutes of observation.  The data from the second night shows a source with a baseline magnitude of $g\sim20$ that dips dramatically, twice: once by 2 magnitudes and a second time by 1.5 magnitudes, each occurring in the space of a few minutes.  Visual inspection of the first night of data revealed no signs of contamination by non-astrophysical sources. The analysis of the data from the second night, shown in the bottom panel of Figure \ref{fading}, reveals that DWF102920.187-355700.211 and the dimmer stars in the vicinity all become very faint; clouds passing over this region of the sky would account for the apparent dimming of the source during the second night, if the clouds passed over this region of the sky and not the region containing reference stars for the field.  We believe that DWF102920.187-355700.211 was displaying some genuine transient phenomena on the first night of observation before reaching a quiescent phase in the second and third nights.

In Figure \ref{inflex}, we present two light curves that showcase when the FTF can identify a fast evolving transient in the DWF ``real-time'' data stream, and how quickly astronomers can trigger other resources.  The left panel shows the source DWF040623.456-550041.171 around $g\sim17$ before dropping by 0.7 magnitudes over a 10 minute period.  For a source like this, the FTF algorithm would alert astronomers within the first few data points (within 10 minutes in the case) after the light curve deviates from a flat position.  The right panel of Figure \ref{inflex} shows the light curve of DWF102613.233-350150.332 rising by about 0.8 magnitudes in 40 minutes, before undergoing a seemingly exponential decay over the remainder of the observations.  This source would be identified as a potential transient after the first 5 data points, due to the steep nature of its increasing brightness.  Furthermore, the FTF algorithm would identify an inflection point within 10 minutes of the object's drop in brightness, notifying astronomers of a change in the behavior of this object.  

\subsection{Transient Misidentification}
\label{confounding}

In this section we present a sample of light curves that were identified as possible transients by the FTF algorithm, but after further analysis, were determined to be bogus.  The most common type of light curve that confounded the FTF algorithm were those involving an astronomical source interacting with the edge of one of the 62 science CCDs that make up the DECam detector \citep{decam}, pictured in Figure \ref{survey_table} under the label DES (Dark Energy Survey); the number of CCDs increases the chance for edge interactions.  As the source moves onto or off of a CCD the light curve can show a peak or a dip not unlike that mimicking a fast rising or fading transient. This effect was exacerbated by early DWF observational strategies employing dithering routine (e.g., the first run on the 4-hour field analyzed in this paper); dithering patterns are no longer favored by DWF, in part for this reason.  This issue can be remedied by ignoring data collected near the edge of a detector. This information is not always available in cataloged data sets, but can be easily identified using software analyzing the dimensions of the science image (i.e. is it square?) and by machine learning algorithms.

\begin{figure}
    \centering
    \plottwo{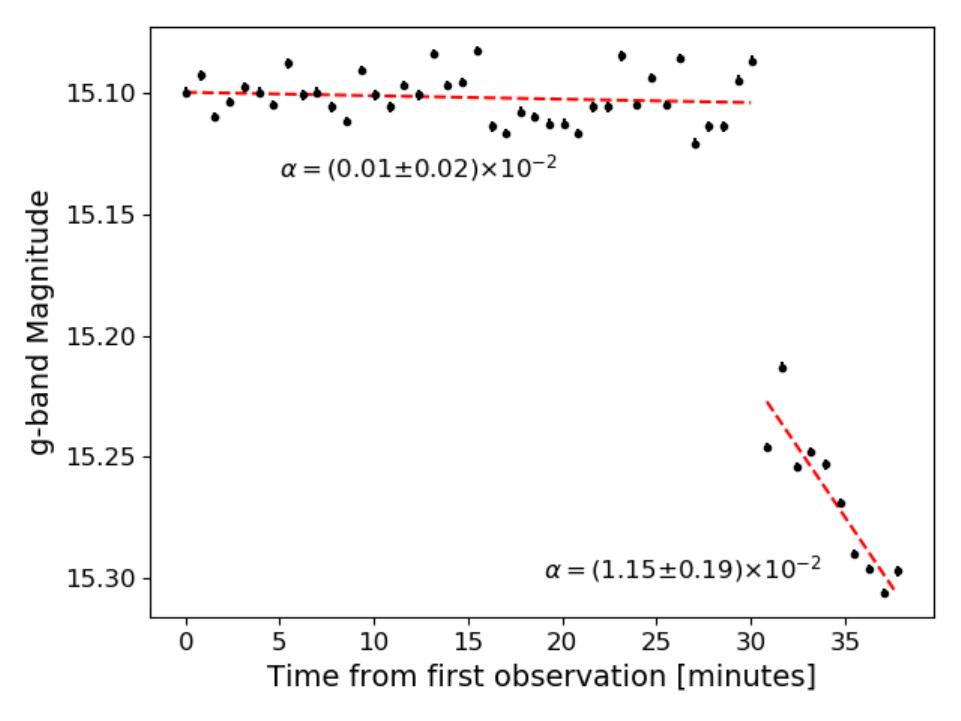}{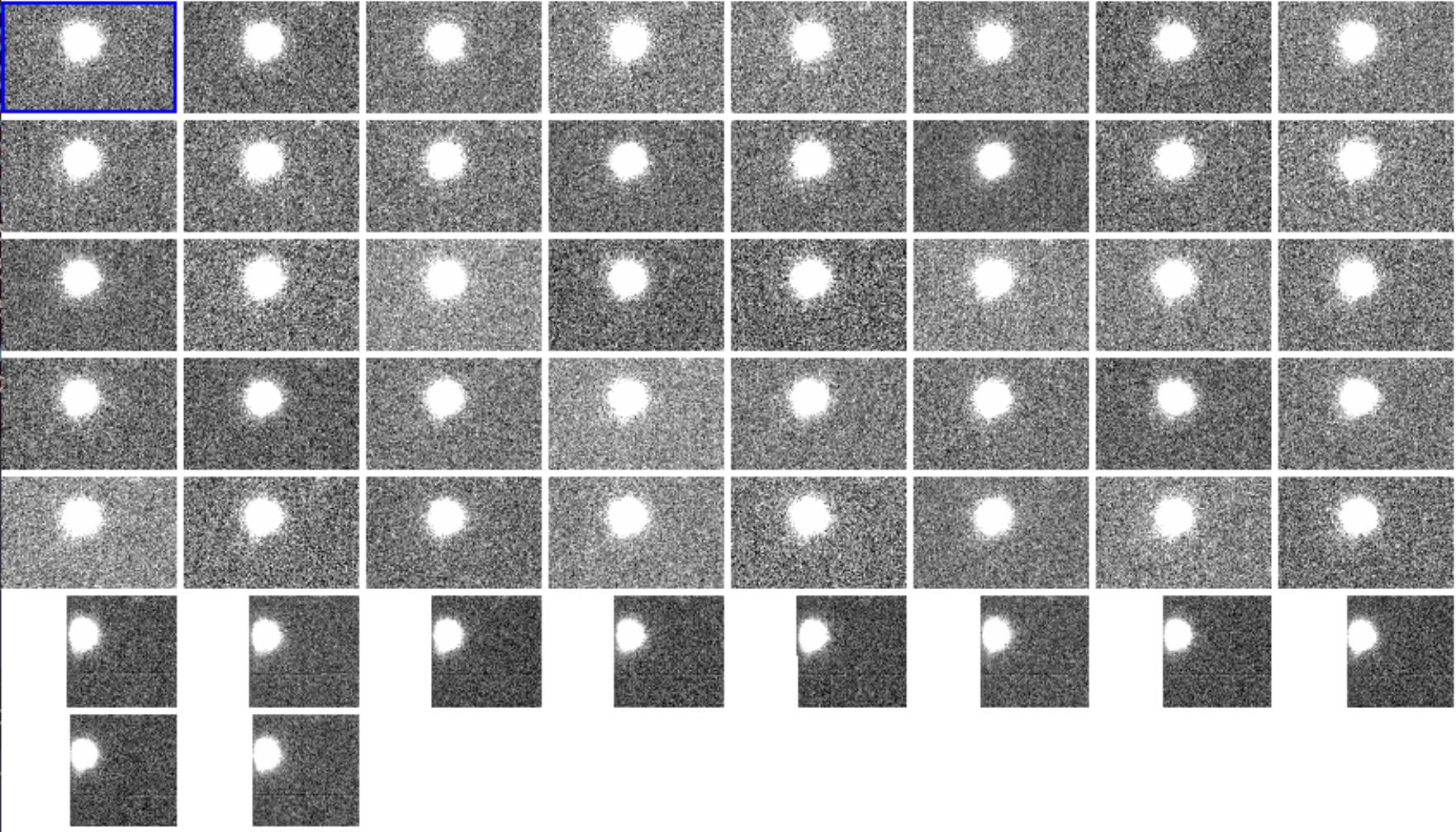}
    \caption{Left: Light curve of DWF040903.800-554603.567 that was flagged by the FTF algorithm as a potential transient, due to the fast 0.1 magnitude drop, and the slower, further 0.1 magnitude decay.  Right: Image cutouts centered on the position of DWF040903.800-554603.567 on the sky.  These image cutouts show the source moving off the edge of the CCD, causing the decreasing light curve.}
    \label{edge_example}
\end{figure}

In Figure \ref{edge_example}, we present an example of an astronomical source DWF040903.800-554603.567 appearing to exhibit transient behavior.  In the left panel of Figure \ref{edge_example}, the light curve dims by $>0.1$ magnitude in one minute before continuing to decay over the next five minutes.  Upon visual inspection of the fits images (in the right panel of Figure \ref{edge_example}, it is clear that the telescope shifted slightly, placing the source on the edge of the detector, and afterwards, the sources is slowly moving out of frame.

\begin{figure}
    \centering
    \plottwo{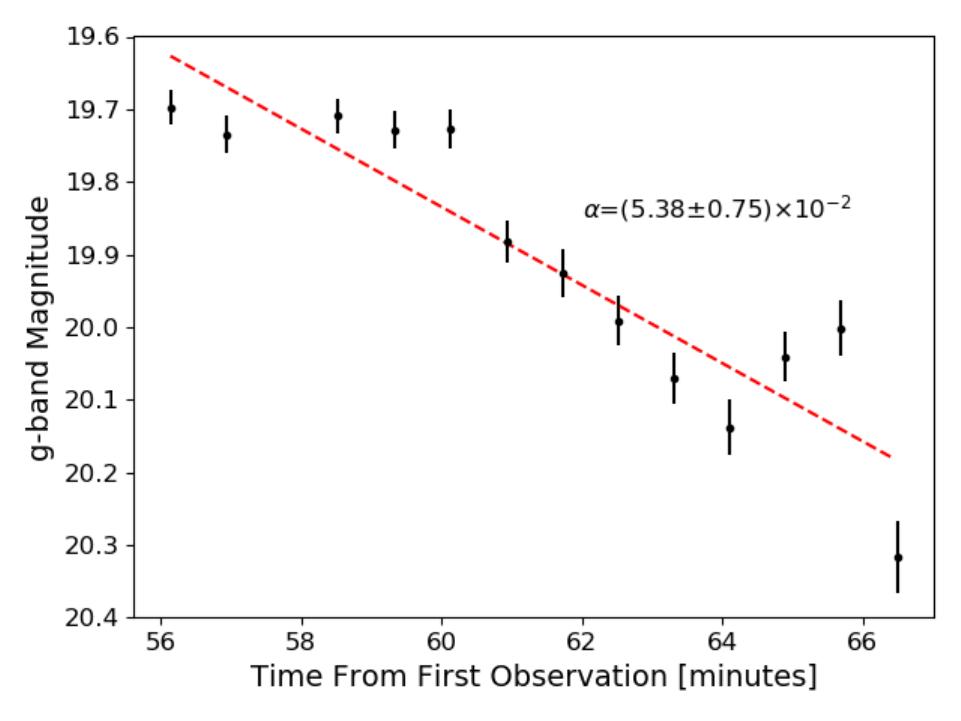}{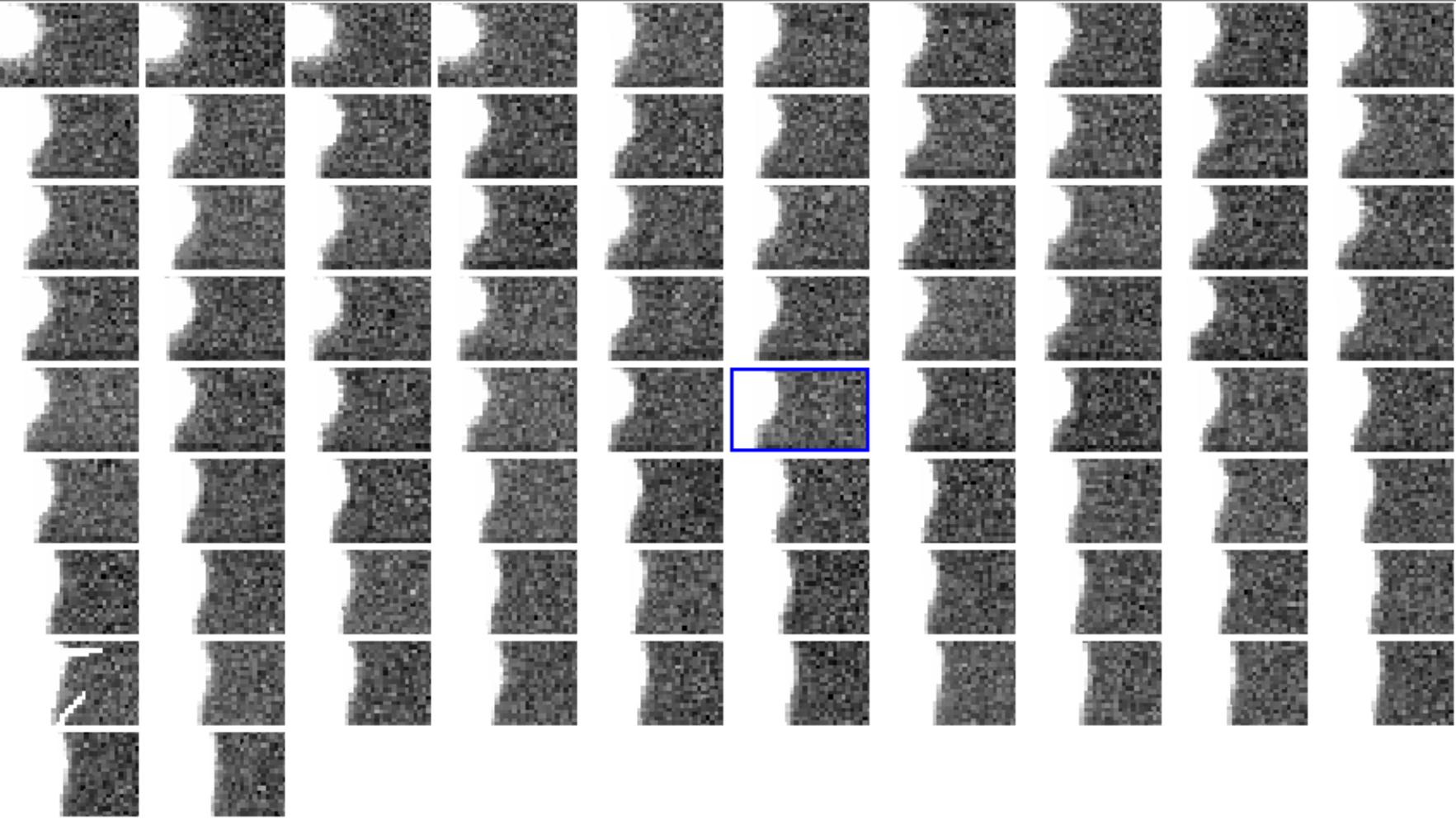}
    \caption{Left: Light curve of DWF040748.870-541956.717 that was flagged by the FTF algorithm as a potential transient.  This source seems to appear at $g\sim19.7$ before rapidly fading by $\sim$0.6 magnitudes in 7 minutes.  Right:  Image cutouts centered on the position of DWF040748.870-541956.717 on the sky.  
    Upon further inspection, it appears that a bright source is on the edge of the CCD; as it moves further off the edge, a small section of the star is still visible.  This small bit of flux is assigned to coordinates still in the field of view of this CCD, and a new source is generated in the catalog.  As the source continues to move off the CCD, the measured flux decreases.}
    
    \label{edge2}
\end{figure}

In Figure \ref{edge2}, we present a light curve that was 
misidentified as a transient due to edge effects, but for a 
slightly different reason than that shown in Figure 
\ref{edge_example}.  In the left panel of Figure \ref{edge2}, 
we present the light curve of DWF040748.870-541956.717 that 
appears at a magnitude of $g\approx19.7$ out of a $5\sigma$ 
background upper limit of $g<22.5$.  The source then proceeds 
to decay by $0.6$ magnitudes over the course of about 7 
minutes.  Upon inspection of the images, shown in the right 
panel of Figure \ref{edge2}, a bright source is shown moving 
out of frame.  We suspect that the cause of the appearance of 
the source at $g\approx19.7$ more than 50 minutes after the 
observation of the field began is the following: 1) a bright 
star was present in the field at coordinates slightly offset
from those of DWF040748.870-541956.717, 2) this bright source 
began to move out of frame, 3) eventually the centroid of the 
star is out of frame, but light from the star is still being 
detected; the NOAO pipeline then identifies a new source using 
coordinates in frame 4) as the source continues to move out of 
frame, the brightness of the object continues to decrease.

\section{Conclusions and Future work} \label{conclusion}

The Deeper, Wider, Faster program is unique in terms of its depth ($g\sim23$ per image) and its short cadence ($\sim$ 1 minute) when compared to other transient surveys, occupying a parameter space with a distinct lack of coverage \citep[e.g.,][Figure 6]{dwf_parameterspace}.  
In addition to its depth and cadence, DWF offers a new way to explore transient phenomena due to the simultaneous wide-field multi-wavelength observations performed across the entire electromagnetic spectrum. Identification of transient phenomena in transient surveys has heavily relied on the imperfect science of image subtraction.  Image subtraction is necessary in some cases, such as the identification of a transient within a bright host galaxy.  Identification of transients via light curve analysis, can be done independently from image subtraction, or in concert with image subtraction techniques. Light curve analysis can identify variable objects with small changes in brightness that might be missed in an image subtraction, for example exoplanet transits.  In addition, the rudimentary classification of transient phenomena requires analysis of the light curves of these objects, with more refined classifications relying heavily on a spectral analysis of the object.

In this work, we present the Fast Transient Finding (FTF) algorithm, capable of identifying transient phenomena both independently of image subtraction (e.g., ``Post-run Data'' in Section \ref{realtimeresults}, and in tandem with an image subtraction algorithm (e.g., ``Real-Time Data'' in Section \ref{realtimeresults} and the $*$ fields in Table \ref{results_table}), on the DWF data stream light curves.  We focused on identifying fast transients (e.g., explosive phenomena) in this paper, but also demonstrate how the FTF algorithm can be customized to find other kinds transients and variables.  

This type of algorithm occupies a unique space within the transient detection landscape.  Most currently operating optical surveys do not detect intra-night variability, and as such, miss the opportunity to alert the community for possible follow-up on fast evolving transients such as GRB and FRB counterparts. 



We see the work in this paper as the first step towards implementation of real time transient classification.  We will first identify potential transients using the FTF algorithm.  Next we will combine the multi-wavelength data sets obtained by the DWF for sources of interest.  We will either extract features from this combined multi-frequency data set or run a deep learning classification algorithm in real-time (Cucchiara et al. in preparation).

\begin{figure}
    \centering
    \plotone{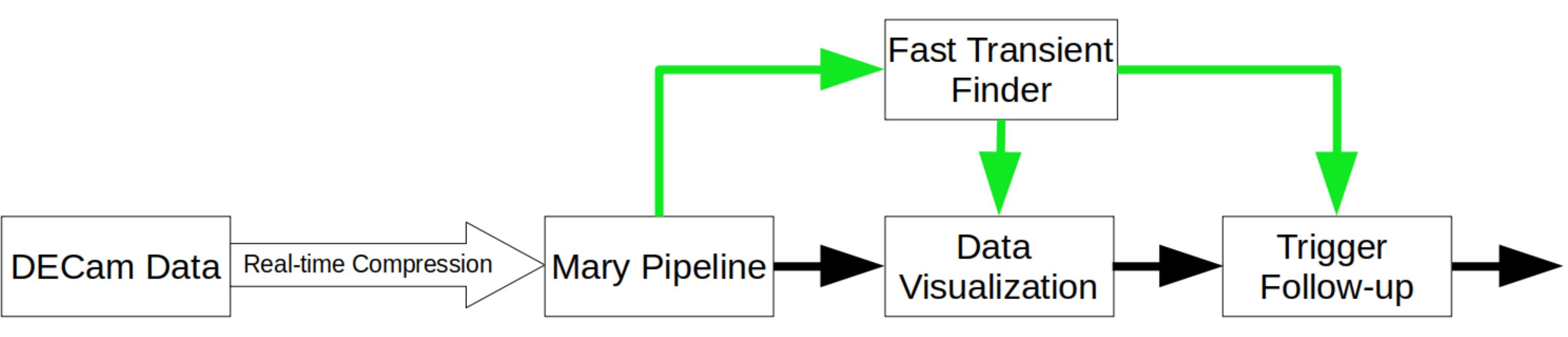}
    \caption{Current data flow for incorporation of the FTF algorithm into the DWF pipeline.  The green arrows represent the new data flows described in this paper.  Data from DECam are compressed \citep{dwf_data_comp} and sent to Australia for analysis.  The images are processed by the Mary Pipeline \citep{mary} and image subtractions are performed.  Light curves of sources flagged through image subtraction are fed to the FTF algorithm.  The light curves are analyzed and the results are fed to the Data Visualization \citep[e.g.,][]{dwf_visualization}, or used directly to trigger follow-up observations.}
    \label{current_flow}
\end{figure}

The FTF algorithm will be incorporated into the DWF pipeline and deployed on the next DWF run, as shown in Figure \ref{current_flow}.  In its first iterations the algorithm will be working off of the light curves generated by the image subtractions performed by the Mary pipeline \citep{mary}.  When a source is first identified as a candidate by image subtraction, a light curve will begin to be populated for that source.  If the slope of the light curve of that source is above some threshold, which we can select manually for specific sources (very high for flare stars or slightly lower for slower evolving transients) or automatically using a statistical measure (e.g. Figure \ref{slopehistograms}), then that source will be identified as a potential fast transient candidate.  Candidates from the image subtraction are provided to human observers using interactive visualization tools.  We will give priority to sources that are flagged as potential transients by the FTF algorithm, as these sources are both image-subtracted candidates and FTF candidates.  As more data is generated, sources with more inflection points will drop out of the FTF candidate list.  We can trigger followup of image subtracted and FTF candidates to classify these sources in real-time (e.g. with detailed spectra).

Due to the general nature of the FTF algorithm, we will look to apply it to other data sets, both propreitary and publicly available.  In particular, some authors of this paper are members of the Rubin Science Collaboration or are Rubin Observatory Data Preview 0.1 (DP0\footnote{https://dp0-1.lsst.io/dp0-delegate-resources/index.html}) Delegates, and have early access to the Rubin Science Platform.  We plan to test the FTF algorithm on the DP0 data set and refine our algorithm before Rubin Observatory comes fully online in 2023.

\acknowledgments
RS and AC are supported by the NSF Grant AST\# 1831682.

This project used data obtained with the Dark Energy Camera (DECam), which was constructed by the Dark Energy Survey (DES) collaboration. Funding for the DES Projects has been provided by the US Department of Energy, the US National Science Foundation, the Ministry of Science and Education of Spain, the Science and Technology Facilities Council of the United Kingdom, the Higher Education Funding Council for England, the National Center for Supercomputing Applications at the University of Illinois at Urbana-Champaign, the Kavli Institute for Cosmological Physics at the University of Chicago, Center for Cosmology and Astro-Particle Physics at the Ohio State University, the Mitchell Institute for Fundamental Physics and Astronomy at Texas A$\&$M University, Financiadora de Estudos e Projetos, Fundação Carlos Chagas Filho de Amparo à Pesquisa do Estado do Rio de Janeiro, Conselho Nacional de Desenvolvimento Científico e Tecnológico and the Ministério da Ciência, Tecnologia e Inovação, the Deutsche Forschungsgemeinschaft and the Collaborating Institutions in the Dark Energy Survey.

The Collaborating Institutions are Argonne National Laboratory, the University of California at Santa Cruz, the University of Cambridge, Centro de Investigaciones Enérgeticas, Medioambientales y Tecnológicas–Madrid, the University of Chicago, University College London, the DES-Brazil Consortium, the University of Edinburgh, the Eidgenössische Technische Hochschule (ETH) Zürich, Fermi National Accelerator Laboratory, the University of Illinois at Urbana-Champaign, the Institut de Ciències de l’Espai (IEEC/CSIC), the Institut de Física d’Altes Energies, Lawrence Berkeley National Laboratory, the Ludwig-Maximilians Universität München and the associated Excellence Cluster Universe, the University of Michigan, NSF’s NOIRLab, the University of Nottingham, the Ohio State University, the OzDES Membership Consortium, the University of Pennsylvania, the University of Portsmouth, SLAC National Accelerator Laboratory, Stanford University, the University of Sussex, and Texas A$\&$M University.


\bibliographystyle{aasjournal}
\bibliography{DWF_bib}{}



\end{document}